\documentstyle[12pt]{article}
\newtheorem{theorem}{Theorem}
\newtheorem{itlemma}{Lemma}[section]
\newtheorem{itproposition}[itlemma]{Proposition}
\newtheorem{itcorollary}[itlemma]{Corollary}
\newtheorem{itremark}[itlemma]{Remark}
\newtheorem{itremarks}[itlemma]{Remarks}
\newtheorem{itdefinition}[itlemma]{Definition}
\newtheorem{itexample}[itlemma]{Example}

\newenvironment{lemma}{\begin{itlemma}\rm}{\end{itlemma}} 
\newenvironment{remark}{\begin{itremark}\rm}{\end{itremark}} 
\newenvironment{remarks}{\begin{itremarks} \rm}{\end{itremarks}}
\newenvironment{corollary}{\begin{itcorollary}\rm}{\end{itcorollary}}
\newenvironment{proposition}{\begin{itproposition}\rm}{\end{itproposition}}
\newenvironment{definition}{\begin{itdefinition}\rm}{\end{itdefinition}}
\newenvironment{example}{\begin{itexample}\rm}{\end{itexample}}
\newenvironment{fact}{\noindent {\em Fact}. \ \ }{\hfill \medskip}
\newenvironment{proof}{\noindent {\em Proof}.\ \
}{\hspace*{\fill}$\Box$\medskip}
\newenvironment{claim}{\noindent {\em Claim}. \ \ }{\hfill \medskip}
\newcommand{\be}[1]{\begin{equation}\label{#1}}
\newcommand{\ee}{\end{equation}}
\newcommand{\bl}[1]{\begin{lemma}\label{#1}}
\newcommand{\br}[1]{\begin{remark}\label{#1}}
\newcommand{\brs}[1]{\begin{remarks}\label{#1}}
\newcommand{\bt}[1]{\begin{theorem}\label{#1}}
\newcommand{\bd}[1]{\begin{definition}\label{#1}}
\newcommand{\bp}[1]{\begin{proposition}\label{#1}}
\newcommand{\bc}[1]{\begin{corollary}\label{#1}}
\newcommand{\bfact}[1]{\begin{fact}\label{#1}}
\newcommand{\bex}[1]{\begin{example}\label{#1}}
\newcommand{\ec}{\end{corollary}}
\newcommand{\efact}{\end{fact}}
\newcommand{\eex}{\end{example}}
\newcommand{\el}{\end{lemma}}
\newcommand{\er}{\end{remark}}
\newcommand{\ers}{\end{remarks}}
\newcommand{\et}{\end{theorem}}
\newcommand{\ed}{\end{definition}}
\newcommand{\ep}{\end{proposition}}
\newcommand{\epr}{\end{proof}}
\newcommand{\bpr}{\begin{proof}}
\newcommand{\bcl}{\begin{claim}}
\newcommand{\ecl}{\end{claim}}

\newcommand{\bi}{\begin{itemize}}
\newcommand{\ei}{\end{itemize}}
\newcommand{\ben}{\begin{enumerate}}
\newcommand{\een}{\end{enumerate}}
\newcommand{\text}[1]{\hbox{\rm \ #1\ \/}}

\newcommand{\RR}{\mbox{${\rm \:  R\!\!\!\! I
\;\;}$}}

\newcommand{\vs}{\vspace{0.25cm}}
\newcommand{\qed}{\hfill $\Box$ \vskip 2ex}

\newcommand{\r}{{{\cal R}}}

\begin{document}

\noindent \hfill
\begin{center}
{\Large {On Quantum State Observability and Measurement}}
\end{center}

\bigskip

\begin{center}

\vs {Domenico D'Alessandro \\ Department of Mathematics\\
Iowa State University \\ Ames, IA 50011,  USA\\ Tel. (+1) 515 294
8130\\ email: daless@iastate.edu}

\end{center}

\vspace{0.5cm}

\begin{abstract}

We consider the problem of determining the state of
a quantum system given one or more readings of the 
expectation value of an    observable. The 
system is assumed to be a finite dimensional 
quantum control system for  which we can influence 
the dynamics  by generating all the
 unitary evolutions in a Lie group. We investigate
to what extent, by an appropriate sequence  of evolutions 
and measurements,  we can obtain information on the initial state of
the system. We present  a system theoretic viewpoint of
this problem in that we 
study the {\it observability} of the  system. In this
context,  we characterize the equivalence 
classes of indistinguishable states and 
propose algorithms for state identification.   

\end{abstract}

\section{Introduction}

Given a control system 
\be{generale}
\dot x=f(t,x,u), 
\ee
where $u$ represents the control function, $x$ the state varying on a
manifold $M$,  with output $y=y(x)$, denote by $x(t,u,x_0)$ the
solution of (\ref{generale}) with control $u$, initial condition $x_0$, 
at time $t$. Two states $x_{01}$ and $x_{02}$ 
are said to be {\it indistinguishable} (see e.g. \cite{Sontag})  if, 
 for every control $u$ and
every time $t$,  we have $y(x(t,u,x_{01}))= y(x(t,u,x_{02}))$. A system is
said to be {\it observable} if no two states in $M$ are indistinguishable.

\vs

In this paper, we are interested 
in the observability properties of
 quantum control systems whose 
dynamics are described by the
Liouville's  equation for the 
density matrix $\rho$ (see e.g. \cite{Blum}),   
\be{scro}
i   \frac{d}{dt} \rho  = [H(u(t)),  \rho].    
\ee    
We shall restrict ourselves to the finite dimensional case where $\rho$
is an $n \times n$ matrix. The {\it Hamiltonian} $H(u(t))$  is  an
$n \times n$ Hermitian matrix, 
in general, function of one or more control functions
$u(t)$. We have assumed here and will 
assume in the rest of the paper that 
we are dealing with closed (noninteracting with the  environment 
if not through the control functions and during 
the measurement process) quantum system. We 
assume that we perform a measurement of the mean value of an observable, 
represented by a Hermitian matrix $S$.  In this case the output $y$  
 takes the form 
\be{firstmeas}
y=Tr(S \rho). 
\ee
Since $Tr(\rho) \equiv 1$, it will be convenient to replace $\rho$ with
the traceless matrix $\rho-\frac{1}{n}I_{n \times n}$ and $S$ with the
traceless matrix $S-\frac{Tr(S)}{n}I_{n \times n}$. This has the
effect of `shifting' the value of the output by a constant $Tr(S)$ value
which does not play any role in the indistinguishability
considerations that will follow. Therefore we will set 
in the following $Tr(\rho)=0$ and 
$Tr(S)=0$. 
The solution of (\ref{scro})
varies as  
\be{evoluden}  
\rho(t)=X(t) \rho(0) X^*(t),    
\ee
\vs
with $X$ solution of the
Schr\"odinger equation,  
\be{bassys}
\dot X(t)=-i H(u(t)) X,  \qquad X(0)=I_{n \times n}.    
\ee
From known results in the theory of quantum control 
(see e.g. \cite{Tarn}, \cite{sussJ}, \cite{Ramak},   and see \cite{MIKOnew}, 
\cite{Jurde} for 
the non-bilinear case), $X$ can be driven to every value 
in the Lie group $e^{\cal L}$ corresponding to the Lie algebra $\cal L$
generated by $
span_{u \in {\cal U}} \{ iH(u) \}$ where $\cal U$ denotes the 
set of possible values for the control. With 
initial condition $\rho(0)$, 
Hermitian and with trace zero, 
the density matrix $\rho$
can attain all the values in the orbit 
\be{Orbita}
{\cal O}:= \{ X \rho(0) X^*| X \in e^{\cal L} \}. 
\ee  

\vs


A study of the observability of control systems involves two
main things. First, one would like to collect, in equivalence classes,  
initial  states that cannot be
distinguished by varying the control and measuring the output 
(see the next two sections for definitions in our case). 
Second, one would like to have  methods to infer the 
equivalence class of the initial state from appropriate sequences of 
measurements and evolutions. We consider these problems for 
quantum control systems in this
paper. 

\vs 

The question  of determination of the state of a quantum system
from measurements is at the heart of quantum mechanics and it was
already discussed by Pauli in \cite{Pauli}. Several
contributions have appeared in recent years and a discussion of the
problem in general terms can be found in \cite{Busch}, where, like in
the present paper,
the problem of determination of the {\it initial} state 
(as opposed to the {\it current} state) was described. We present in
this paper a 
treatment of this topic from a system theoretic view point. In this
context, our study is closely related to other studies of the
observability of nonlinear systems \cite{HK}, \cite{Is}, \cite{NV} (see
also \cite{Lieobs} for  systems varying
on Lie groups). However we consider here a specific model for which we
can obtain more complete results. Moreover  a
new element appears in the treatment of quantum systems,  that is the
transformation of the state as a result of each measurement. This can
take different forms according to the type of measurement considered 
(see e.g. \cite{BK}, \cite{BLM}, \cite{Davies}). We shall mainly consider 
 the case
of Von Neumann measurement   \cite{sakurai},
\cite{vonNeu} and discuss extensions to other cases.  

\vs 

The paper is organized as follows. In Section 2 we define and 
describe the set of states that cannot be distinguished in one  
measurement. Then, we generalize in Section 3 to states that cannot be
distinguished in multiple measurements. The determination of the state
from one or more measurements is
discussed in Section 4. Conclusions are given  in Section 5. 

\vs

\section{Indistinguishability 
and observability with  a single 
measurement}

In the following, $S$ is  the traceless 
Hermitian matrix representing the 
observable and  $\rho(t,u,\rho_0)$ 
is  the solution of (\ref{scro}) at time $t$, with 
initial condition equal to  $\rho_0$, and control $u$.

\vs 

\noindent {\bf Definition 1.} 
Two states $\rho_1$ and $\rho_2$ are 
{\it indistinguishable in one step} 
if,  for every control
function(s) $u$ and every $t$, we have 
\be{p1}
Tr(S\rho(t,u,\rho_1))=Tr(S\rho(t,u,\rho_2)).
\ee

\vs 

The definition asserts that two states $\rho_1$ and $\rho_2$ 
are indistinguishable if 
there is no admissible experiment involving
only one measurement which would give different results with initial
states $\rho_1$ and $\rho_2$. It is clear that indistinguishability in
one step is an equivalence relation. The set of possible
values for the density matrix  will be denoted by $\cal R$.  It is a 
 convex subset  of the vector space 
of $ n \times n$ Hermitian matrices (with zero trace),   $i
su(n)$ and, in general,  the vector space spanned by the
elements of $\cal R$ is the same as $i(su(n))$. 
The elements of $\cal R$ are parametrized by
$n^2-1=\dim  su(n)$ parameters \footnote{In the presence of special
symmetries, a parametrization with fewer parameters can be given (see 
 \cite{GS} for an example).} 

\vs 

\noindent {\bf Definition 2.} The 
system is {\it observable in one step}
 if $\rho_1$ and $\rho_2$ $\in \r$  
 are indistinguishable in one step only when   $\rho_1=\rho_2$.

\vs

Instrumental in the characterization of classes of indistinguishable 
states is the vector space of $n \times n$ 
skew-Hermitian matrices,
\be{cool}
{\cal V} : = \oplus_{k=0}^\infty ad_{\cal L}^k iS.  
\ee
Here  $ad_{\cal L}^k iS$ is 
the space obtained by taking $k$ Lie brackets of
$iS$ with elements in the Lie algebra $\cal L$.
We shall call $\cal V$, {\it Observability
Space}. If $B_1,...,B_m$ is a set of generators 
of the Lie algebra $\cal L$, it follows from an application of the
Jacobi identity (see Appendix A) that  the
observability space  $\cal V$ is spanned by the
matrices\footnote{$ad_R^kT:=[R,[R,...[R,T]]]$ where the Lie bracket is
taken $k$ times.} 
\be{mat}
ad_{B_{j_1}}^{k_1} ad_{B_{j_2}}^{k_2} 
\cdot \cdot \cdot ad_{B_{j_r}}^{k_r} iS, 
\ee
with  $k_1,...,k_r \geq 0$, and $\{ j_1,...,j_r \} \in
\{1,...,m\}$.  $\cal V$ is the smallest subspace of $su(n)$ 
{\it stable} under
$\cal L$ 
\footnote{$[{\cal L}, {\cal V}] \subseteq \cal V$.} and
containing $iS$.  $\cal V$ might not be a 
Lie Algebra, however it is always a subspace  of the Lie Algebra
generated by $iS,B_1,...,B_m$ and therefore a subspace of $su(n)$.  
Therefore its dimension is bounded by $\dim  su(n)=n^2-1$. Notice that
$\cal V$ can be  calculated with an algorithm that, at each step,
calculates the matrices of `depth' $d+1$ from the matrices of depth $d$, 
where the depth is the number of Lie brackets performed namely $k_1+k_2+
\cdot \cdot \cdot + k_r$ in (\ref{mat}). The algorithm starts with the
matrix $iS$, which has depth $0$, and ends when the dimension reaches $n^2-1$
or there is no increment in the dimension. 
By finite dimensionality,  there is always a finite $\bar k$ such that 
\be{cool2}
{\cal V} = \oplus_{k=0}^{\bar k} ad_{\cal L}^k iS. 
\ee

\vs
 
We have the following result that relates the partition  of the state
space into classes of indistinguishable states with the properties of the
observability space $\cal V$. 

\vs

\noindent{\bf Theorem 1.} {\it The following three conditions are 
equivalent 

\begin{enumerate}

\item The states $\rho_1$ and $\rho_2$ are indistinguishable in one step.

\item For every $X \in e^{\cal L}$, 
\be{condfra}
Tr(X^* S X \rho_1)=Tr(X^* S X \rho_2). 
\ee

\item For every $F \in {\cal V}$, 
\be{condit}
Tr (F \rho_1)=Tr(F \rho_2). 
\ee
\end{enumerate}

}

\vs

\noindent {\bf Proof.} The equivalence between conditions $1$ and $2$
simply follows from the fact that the set of values obtainable for
$\rho$  starting from $\rho(0)=\rho_{1,2}$ 
is described in (\ref{Orbita}), and 
from elementary properties of the trace.  

Now assume (\ref{condfra}) holds and choose $k$ matrices $R_1,...,R_k$
(not necessarily all different) 
in   $\cal L$. 
Then, for every $k-$ple of real numbers
$t_1,...,t_k$ we have 
\be{rt}
Tr(e^{-R_1 t_1} \cdot \cdot \cdot e^{-R_k t_k} i S e^{R_k t_k} 
\cdot \cdot \cdot e^{R_1 t_1} 
\rho_1)=
Tr(e^{-R_1 t_1} \cdot \cdot \cdot e^{-R_k t_k} iS e^{R_k t_k} 
\cdot \cdot \cdot e^{R_1 t_1} \rho_2).  
\ee 
Calculating the derivative,  
$
\frac{\partial^k}{\partial t_1 \partial t_2 \cdot \cdot \cdot \partial
t_k}_{t_1=t_2=\cdot \cdot \cdot = t_k=0}, 
$ of both sides 
we obtain 
\be{what}
Tr(ad_{R_1}ad_{R_2} \cdot \cdot \cdot ad_{R_k} iS \rho_1)= 
Tr(ad_{R_1}ad_{R_2} \cdot \cdot \cdot ad_{R_k} iS \rho_2), 
\ee
which proves Condition 3, 
since $k$ and $R_j$, $j=1,...k$, are not
specified. To prove that Condition 3  implies Condition 1, 
let $F_1,....,F_s$ be a basis of $\cal V$ with  $F_1=iS$. 
Then we have, using (\ref{scro}),   
\be{derivata}
\frac{d}{dt} Tr(F_j \rho(t, u,\rho_{1,2}))=
\sum_{k=1}^s a_{j,k}(t) Tr(F_k \rho(t, u,\rho_{1,2})),  
\ee  
for some (time varying)  coefficients $a_{j,k}(t)$ depending on the
control $u$.  Therefore we have that 
$Tr(F_j \rho(t, u,\rho_{1}))$ and  
$Tr(F_j \rho(t, u,\rho_{2}))$, satisfy the same (linear) 
system of differential equations and since the initial conditions are the same, then 
\be{equalitad}
Tr(F_j \rho(t, u ,\rho_{1}))= 
Tr(F_j \rho(t, u ,\rho_{2})), \qquad j=1,...,s. 
\ee
In particular, we have,  
\be{uscite}
Tr(S \rho(t, u,\rho_{1}))= 
Tr(S \rho(t, u,\rho_{2})). 
\ee
Therefore the two states are indistinguishable. 
\qed

\vs

The inner product $<\cdot,\cdot>$ in $su(n)$ is defined as 
$<A,B>=Tr(AB^*)$. Theorem 1 states that two matrices in $\cal R$ 
are indistinguishable if and only if they differ by
an element in ${\cal V}^\perp$. Therefore we can state the following
criterion of observability in one step which is a consequence of Theorem
1. 

\vs

\noindent {\bf Theorem 2.} {\it System (\ref{scro}) is 
observable in one step if and
only if one of the  following equivalent 
conditions are verified

\begin{enumerate}

\item 
\be{cond1t}
span_{X \in e^{\cal L}}X^*iSX=su(n), 
\ee

\item
\be{cond3}
{\cal V}=su(n). 
\ee

\end{enumerate}} 

\vs 

\noindent {\bf Remark:} The notion of observability is closely 
related to the notion of {\it informational completeness} of observables
as treated for example in \cite{HS}. A set of observables $\cal B$ is
called  informationally complete if $Tr(B\rho_1)=Tr(B\rho_2)$ for every
$B \in \cal B$ implies $\rho_1=\rho_2$. From condition $2$ of Theorem 1
and the definition of observability, we can say that a system is observable
if and only if the set of operators $\{X^*S X| X \in e^{\cal L}\}$ is
informationally complete. 

\vs

\subsection{Relation between controllability 
and observability in one step}

If ${\cal L}=su(n)$, namely the system is
{\it operator controllable} \cite{confraIEEE}, 
and $S \not=0$,  then it is also observable. 
In fact,  in this case,  we have 
\be{cond1}
span_{X \in e^{\cal L}}X^*iSX=span_{X \in SU(n)}  X^*iS X=su(n). 
\ee
To verify this we can more easily verify condition (\ref{cond3}). 
Since $\cal V$ is a nonzero 
 ideal of $su(n)$ and $su(n)$ is a simple Lie algebra, $\cal V$ must be
equal to $su(n)$. Therefore we have. 

\vs

\noindent {\bf Corollary 3} {\it Controllability along with $S \not=0$ 
implies observability in one step.}

\vs 

The converse of Corollary 3 is not true not only because we may  have the
equality 
\be{orbite}
\{X^*SX|X \in e^{\cal L} \} = \{X^*SX|X \in SU(n) \},  
\ee
even though ${\cal L} \not=  su(n)$ \cite{confraIEEE}, \cite{Schirmer}    
but also because we may  have
(\ref{cond1t}) (\ref{cond3})  even though 
(\ref{orbite}) is not verified. A simple
example of this can be found  already in the $n=2$ case by taking 
\be{esempio}
iS=\pmatrix{ i & 1 \cr -1 & -i}, \qquad {\cal L}= span\{ \pmatrix{ 0 & 1
\cr -1 & 0} \}. 
\ee

\subsection{First order conditions for observability in one step.}

The case of the equality of the orbits in (\ref{orbite}) is 
particularly favorable because we can 
give a different condition of
observability which avoids the calculation of repeated Lie brackets for
$\cal V$ and involves only the calculation of Lie brackets of depth
1. We have the following Proposition.

\noindent {\bf Proposition  4} {\it Assume $S \not=0$. 
The system is observable in one step if one of the
following two equivalent conditions is verified
\begin{enumerate}
\item 
\be{orbite2}
\{X^*SX|X \in e^{\cal L} \}= \{X^*SX|X \in SU(n) \},  
\ee

\item 
\be{micond}
[{\cal L}, iS] = [su(n), iS]. 
\ee
\end{enumerate}
} 

\vs 

From a practical point of view condition (\ref{micond}) may be easier to
verify since it involves calculation of first order Lie brackets
only. The condition tells us that,  by calculating first order Lie
brackets, we can infer the properties of $\cal V$ which is defined
through higher order Lie brackets. If condition (\ref{micond}) is not
verified we may still have observability. 

\vs 

\noindent {\bf Proof.} The equivalence between the 
conditions  (\ref{cond1}) and (\ref{orbite2}) was proven 
in \cite{confraIEEE} although in a different context\footnote{There, 
$S$ was the density matrix and we wanted to give practical conditions to
verify that the set of possible density matrices that can be obtained by
varying $X$ in $e^{\cal L}$ is the same as the largest possible one
namely the one obtained by varying $X \in SU(n)$.  This condition was
called {\it Density Matrix Controllability}.}, therefore we shall not
repeat the proof here. Clearly (\ref{orbite2}) implies
(\ref{cond1t}) with (\ref{cond1}) and (\ref{cond3}) and therefore
observability. \qed

\vs 

Condition (\ref{micond})  can be 
verified by comparing the dimensions of
the two vector spaces. The dimension of $[iS, su(n)]$ can be expressed
in terms of the multiplicity of the eigenvalues 
of $iS$ (recall that $iS$ is not zero and it has zero trace so 
it has at least two distinct 
eigenvalues). We have 
\be{dimensione}
\dim [iS, su(n)]= 2 \sum_{j<k} n_j n_k,   
\ee  
where $n_j$ ($n_k$) is  the multiplicity of the $j-$th ($k-$th)
eigenvalue.

\vs

If $iS$ is known to be in a proper subspace $\cal F$ of $su(n)$
stable under $\cal L$ (e.g. $\cal L$ itself or ${\cal L}^\perp$) 
then we cannot have observability because ${\cal V} \subseteq {\cal F}
\not=su(n)$. 

\vs

\noindent{\bf Example:} Two spin $\frac{1}{2}$ particles are
interacting through Ising interaction and are driven by an
electro-magnetic field in the $x$ direction \cite{Ernst}, 
\cite{EPR}. The magnetic field couples
with one of the spins only and we can detect the magnetization in the
$z$ direction. Denote by  $\sigma_{x,y,z}$  the $x,y,z$ Pauli matrices (see
e.g. \cite{sakurai})
\be{Pauli}
\sigma_x:= \frac{1}{2} \pmatrix{0 & 1 \cr 1 & 0},  
\qquad \sigma_y:=\frac{1}{2} \pmatrix{0 & -i \cr 
i & 0}, \qquad \sigma_z:=\frac{1}{2}\pmatrix{1 & 0 \cr 
0 & -1}, 
\ee
by $\bf 1$ the $2 \times 2$ identity matrix 
 and by $u=u(t)$ the $x$ component of 
magnetic field.   After appropriately 
scaling the parameters involved,  
the Hamiltonian $H$ has the form 
\be{esHamil}
H=\sigma_z \otimes \sigma_z + u(t) \sigma_x \otimes {\bf 1},  
\ee 
and the output matrix $S$ is given by $S=\sigma_z \otimes {\bf 1}+ {\bf 1}
\otimes \sigma_z$. The dynamical Lie algebra $\cal L$ is spanned by 
$i \sigma_z \otimes \sigma_z$, $i {\bf 1} \otimes \sigma_x$ and 
$i \sigma_z \otimes \sigma_y$. We have from formula (\ref{dimensione}) 
$\dim  [iS, su(n)]=8$ while $\dim [iS, {\cal L}]=2$. Therefore the
sufficient criterion of observability of Proposition 4 fails. Moreover since
$iS$ is in  ${\cal L}^\perp$, and ${\cal L}^\perp$ is stable under $\cal
L$, the system is not observable.  

\vs

\subsection{Decomposition of the state space}

It is natural to decompose the state $\rho$ as
$\rho(t)=\rho_1(t)+\rho_2(t)$, with $\rho_1(t) \in \cal V$ and
$\rho_2(t) \in {\cal V}^\perp$, for every $t$. Then we have 
\be{OS1}
\dot \rho_1=-i[H(u), \rho_1], 
\ee
\be{OS2}
\dot \rho_2=-i[H(u), \rho_2], 
\ee
and 
\be{OS3}
Tr(S\rho(t))=Tr(S\rho_1(t)), 
\ee
for every $t$. Therefore if we are interested in the effect on the
output $S$ we can parametrize only the component of $\rho$ in $\cal V$.


\section{Indistinguishability and observability 
with  multiple measurements}

We now generalize the above characterization of states that are
indistinguishable after one measurement to states that are
indistinguishable after $k$ measurements, for general $k$. In 
fact it may happen that,  even if two states give 
the same output function at the first measurement,  
for every control and at every time,  they give 
different values at the second measurement. This
is a consequence of the fact that the first measurement
modifies the state. Modern quantum measurement theory  (see e.g. \cite{BK}, 
\cite{BLM}, \cite{Davies}) has studied ways to model the change in the
state due to measurement as well as ways to integrate the measurement process
in the framework of quantum dynamics. We shall remark at the end of this
section on possible extensions to other types of measurements but will 
consider the
simplest case  where the quantum measurement
postulate \cite{sakurai} \cite{vonNeu} holds. This is also called Von
Neumann (or Von Neumann-L\"uders) measurement.   More precisely, 
rewrite the observable matrix $S$ as 
\be{decom}
S=\sum_{j=1}^n \lambda_j a_ja_j^*:= \sum_{j=1}^n \lambda_j \Pi_j, 
\ee 
where $a_j$ are the orthonormal eigenvectors of $S$, $\Pi_j$, $j=1,...,n$,
are  the
associated projection matrices defined by  $\Pi_j:= a_ja_j^*$ and
$\lambda_j$ are the associated eigenvalues. Define the automorphism  
$\cal P$ in the space of (skew)Hermitian matrices 
\be{calp}
{\cal P}(F)=\sum_{j=1}^n \Pi_j F \Pi_j, 
\ee
which returns the diagonal part of $F$,  
if we are working in a basis where $S$ is diagonal. If 
the state at the time of the measurement is 
$\rho(t,u,\rho_0)$, according to the measurement postulate, 
the state after the measurement is 
${\cal P}(\rho(t,u,\rho_0))$. 
Assume  the experiment  consists 
of an evolution for  time $t_1$ with control
$u_1$, followed by a measurement, followed by an evolution for  time $t_2$
with control $u_2$, followed by a measurement, and so on, up to  
an evolution for  time $t_k$ with control $u_k$.  The $k-$th
measurement,  
at time $t_1+t_2+\cdot \cdot \cdot t_k$,  gives the result 
\be{risultatok}
y_k(t_1,...,t_k,u_1,...,u_k,\rho_0):=
Tr(S \rho(t_k,u_k,{\cal P}(\rho(t_{k-1}, u_{k-1}, {\cal P}(\cdot
\cdot \cdot {\cal P}(\rho(t_1,u_1,\rho_0) )))))). 
\ee
We can extend Definitions 1 and 2 as follows

\vs

\noindent {\bf Definition 3.} Two states  $\rho_1$ and $\rho_2$ are 
{\it indistinguishable in $k$  steps} if for every sequence of control
function(s),  $u_1,u_2,...,u_k$,  
defined in intervals $[0,t_1)$, 
$[0,t_2)$,...,$[0,t_k]$,  we have 
\be{p3333}
y_k(t_1,...,t_k,u_1,...,u_k,\rho_1)
=y_k(t_1,...,t_k,u_1,...,u_k,\rho_2), 
\ee

\vs 

\noindent {\bf Definition 4.} A system is {\it observable 
in $k$ steps} if no two states are indistinguishable in 
$k$ steps. 

\vs

\noindent {\bf Definition 5} Two states are {\it 
indistinguishable} if they are {\it indistinguishable}  
in $k$ steps for every $k \geq 1$. A system is 
said to be {\it observable} if no two states are indistinguishable. 

\vs 

It is convenient to rewrite the output at the $k-$th measurement in
terms of the values  of the evolution operator $X$ in (\ref{bassys})
 at the endpoints of the intervals $[0,t_1)$,...,$[0,t_k]$.  We call
these values of $X$, $X_1$,...,$X_k$. Using (\ref{evoluden}), 
we have 
\be{alternyk}
y_k:=y_k(X_1,...,X_k,\rho_0)=
Tr(SX_k {\cal P} (X_{k-1}{\cal P}
(\cdot \cdot \cdot {\cal P}(X_1 \rho_0  X_1^*) 
\cdot \cdot \cdot)X_{k-1}^*)X_k).  
\ee  
Therefore an alternative definition of 
indistinguishability in $k$ steps can be given, 
that is $\rho_1$ and $\rho_2$ 
are indistinguishable if for every set of values  
$X_1$,...,$X_k$ in $e^{\cal L}$,  
$y_k(X_1,...,X_k,\rho_1)=y_k(X_1,...,X_k,\rho_2)$. 

\vs

We can give conditions of indistinguishability and observability as in
Theorems 1 and 2, by
introducing {\it Generalized  Observability Spaces}. 
More specifically, define the
{\it Observability Space of order $0$}, ${\cal V}_0:=span \{iS \}$, and
 the {\it Observability Space of order $1$}, ${\cal V}_1:={\cal V}$ in
(\ref{cool}). The
{\it Observability Space of Order $k$}, ${\cal V}_k$, is defined
recursively by 
\be{recursspacek}
{\cal V}_k:=\oplus_{j=0}^\infty ad_{\cal L}^j {\cal P}({\cal V}_{k-1}). 
\ee
It is the largest subspace of $su(n)$ containing ${\cal P}({\cal
V}_{k-1})$ and stable under $\cal L$. It also follows 
from a proof analogous to the one in Appendix A that,  
if $B_1,...,B_m$ is a set of generators  
of the Lie algebra $\cal L$, ${\cal V}_k$ is spanned by the
matrices
\be{mat22222}
ad_{B_{j_1}}^{k_1}ad_{B_{j_2}}^{k_2}  
\cdot \cdot \cdot ad_{B_{j_r}}^{k_r} F, 
\ee
with $F \in {\cal P}({\cal V}_{k-1})$, 
$k_1,...,k_r \geq 0$, and $\{ j_2,...,j_r \} \in
\{1,...,m\}$. Notice also that it follows by induction, since ${\cal
V}_0 \subseteq {\cal V}_1$,  that 
\be{incluX}
{\cal V}_{k-1} \subseteq {\cal V}_k,  
\ee
for every $k \geq 1$. 

\vs

\noindent We have the following generalization of Theorem 1. 

\vs

\noindent {\bf Theorem 5.} {\it The following three conditions are 
equivalent

\begin{enumerate}

\item The states $\rho_1$ and $\rho_2$ are indistinguishable in  $k$ steps. 

\item For every $k-$ple $X_1,...,X_k$ with values in $e^{\cal L}$, 
\be{equalk}
Tr(X_1^* {\cal P}(X_2^* {\cal P}(\cdot \cdot \cdot {\cal P}(X_{k-1}^* {\cal
P}( X_k^* S X_k) X_{k-1}) \cdot \cdot \cdot) X_2)X_1 \rho_1)= 
\ee
$$
Tr(X_1^* {\cal P}(X_2^* {\cal P}( \cdot \cdot \cdot {\cal P}(X_{k-1}^* {\cal
P}( X_k^* S X_k) X_{k-1}) \cdot \cdot \cdot) X_2)X_1 \rho_2). 
$$
\item For every $F \in {\cal V}_k$, 
\be{equalkbis}
Tr(F \rho_1) = Tr(F\rho_2). 
\ee
\end{enumerate}

\vs 
}

\vs   

\noindent It follows from (\ref{equalkbis}) and
(\ref{incluX}) that if two states are indistinguishable 
in $k$ steps they are indistinguishable
in $r$ steps for every $r < k$. In other terms if we can distinguish two
states in $r$ steps we can distinguish them in $k>r$ steps as well. 

\vs

\noindent{\bf Proof:} If $\rho_1$ and $\rho_2$ 
are indistinguishable,  then,  for all the  
$X_1,...,X_k$ in $e^{\cal L}$,  we have 
$y_k(X_1,...,X_k,\rho_1)=y_k(X_1,...,X_k,\rho_2)$ in
(\ref{alternyk}). Now notice that, for a general $\rho_0$,
\begin{eqnarray}
Tr(SX_k {\cal P} (X_{k-1}{\cal P}
(\cdot \cdot \cdot {\cal P}(X_1 \rho_0  X_1^*) 
\cdot \cdot \cdot)X_{k-1}^*)X_k^*) =  \\
 Tr(X^*_kSX_k {\cal P} (X_{k-1}{\cal P}
(\cdot \cdot \cdot {\cal P}(X_1 \rho_0  X_1^*) 
\cdot \cdot \cdot)X_{k-1}^*))= \nonumber \\
 Tr({\cal P}(X^*_kSX_k) X_{k-1}{\cal P}
(\cdot \cdot \cdot {\cal P}(X_1 \rho_0  X_1^*) 
\cdot \cdot \cdot)X_{k-1}^*)=\nonumber \\
 Tr(X_{k-1}^*{\cal P}(X^*_kSX_k) X_{k-1}{\cal P}
(\cdot \cdot \cdot {\cal P}(X_1 \rho_0  X_1^*) 
\cdot \cdot \cdot))= \nonumber \\ 
\cdot \nonumber \\
\cdot \nonumber \\
\cdot \nonumber \\
Tr(X_1^* {\cal P}(X_2^* {\cal P}(\cdot \cdot \cdot {\cal P}(X_{k-1}^* {\cal
P}( X_k^* S X_k) X_{k-1}) \cdot \cdot \cdot) X_2)X_1 \rho_0).  \nonumber 
\end{eqnarray}
Using this for $\rho_0=\rho_1$ and $\rho_0=\rho_2$ along with
(\ref{alternyk}) we see that indistinguishability of $\rho_1$ and
$\rho_2$ in $k$ steps implies equation (\ref{equalk}). The proof that
Condition {\it 2.} implies Condition {\it 3.} is exactly analogous to the
corresponding proof in Theorem 1. The proof that Condition {\it 3.} implies
indistinguishability also is a generalization 
of the corresponding proof in 
Theorem 1, with some more elements that we now 
illustrate. Consider a basis $F_j$, $j=1,...,s$, 
of ${\cal V}_k$ and derive a 
differential equation for 
$Tr(F_j \rho(t,u_1,\rho_{1,2}))$. The 
differential equations
corresponding to $\rho_1$ and $\rho_2$ 
are the same with the same
initial conditions, because of the assumption  (\ref{equalkbis}). 
Therefore, in particular, at time $t_1$, we have 
\be{Traccia}
Tr(F_j \rho(t_1,u_1,\rho_1))=Tr(F_j \rho(t_1,u_1,\rho_2)), 
\ee 
for every $F_j \in {\cal V}_k$ and therefore for every 
$F_j \in {\cal P}({\cal V}_{k-1})$. If $\bar F_j$, $j=1,...,\bar s$,  
is a basis of ${\cal V}_{k-1}$, then we have 
\be{tracci2}
Tr({\cal P}(\bar F_j) \rho(t_1, u_1, \rho_1))=
Tr({\cal P}(\bar F_j) \rho(t_1, u_1, \rho_2)), 
\ee
that is 
\be{tracci3}
Tr(\bar F_j {\cal P}(\rho(t_1, u_1, \rho_1)))=
Tr(\bar F_j {\cal P}(\rho(t_1, u_1, \rho_2))).  
\ee
Now, derive a differential equation for the variables 
$Tr(\bar F_j \rho)$, with $\bar F_j$ a basis of ${\cal V}_{k-1}$, 
 on the second interval of length $t_2$ and with control $u_2$. The
function corresponding to $\rho_1$ satisfy the same differential
equation as the function corresponding to $\rho_2$ and since the initial
conditions are the same, from (\ref{tracci3}), we obtain that for every 
$\bar F_j$ in ${\cal V}_{k-1}$
\be{tracci3bis}
Tr(\bar F_j \rho(t_2, u_2, {\cal P}(\rho(t_1, u_1, \rho_1))))=
Tr(\bar F_j \rho(t_2, u_2, {\cal P}(\rho(t_1, u_1, \rho_2)))).  
\ee
This is, in particular,  true for elements of ${\cal P}({\cal
V}_{k-2})$. Proceeding this way,  after $k$ steps,  we obtain the
equalities of outputs $y_k$ in (\ref{risultatok}) for $\rho_0=\rho_1$
and $\rho_0=\rho_2$, for every $k-$tuple $t_1,...,t_k$ and controls
$u_1,...,u_k$, and therefore indistinguishability. \qed 

\vs 

An example of ${\cal V}_1 \not= {\cal V}_2$ is given by 
\be{esempioaggiunto}
S:=\pmatrix{1 & 0 & 0 \cr 0 & -3 & 0 \cr 0 & 0 & 2},  \qquad 
{\cal L}:= span\{ \pmatrix{i & 0 & 2 \cr 0 & -i & 0 \cr -2 & 0 & 0} \}. 
\ee

\vs 

\noindent We also have the following Theorem concerning 
observability. 

\vs 

\noindent {\bf Theorem 6.} {\it System (\ref{scro}) is observable in k
steps
 if and
only if one of the  following equivalent 
conditions is  verified

\begin{enumerate}

\item 
\be{cond1tbis}
span_{X_1,X_2,...,X_k \in e^{\cal L}}
X_1^* {\cal P}(X_2^* {\cal P}( \cdot \cdot \cdot {\cal P}(X_{k-1}^* {\cal
P}( X_k^* iS X_k) X_{k-1}) \cdot \cdot \cdot) X_2)X_1=su(n), 
\ee

\item
\be{cond3bis}
{\cal V}_k=su(n). 
\ee
A system is observable if and only if there exists a $k$ such that one
of the equivalent conditions (\ref{cond1tbis}), (\ref{cond3bis}) is 
verified. 

\end{enumerate}} 

To check observability we only need to verify (\ref{cond3bis}) for a
finite number of $k$'s until we find a $k$ such that ${\cal
V}_{k-1}={\cal V}_k$ or ${\cal V}_k=su(n)$.

\vs

It is obvious that since controllability (${\cal L}=su(n)$) implies
observability in one step it also 
implies observability in $k$ steps for
every $k$. The natural extension of 
the condition (\ref{orbite2}) of Proposition 4 would be 
\be{natext}
\{X_1^* {\cal P}(X_2^* {\cal P}(\cdot \cdot \cdot {\cal P}(X_{k-1}^* {\cal
P}( X_k^* S X_k) X_{k-1}) \cdot \cdot \cdot) X_2)X_1 | X_1,X_2,...,X_k
\in e^{\cal L} \}=
\ee
$$
\{X_1^* {\cal P}(X_2^* {\cal P}(\cdot \cdot \cdot {\cal P}(X_{k-1}^* {\cal
P}( X_k^* S X_k) X_{k-1}) \cdot \cdot \cdot) X_2)X_1 | X_1,X_2,...,X_k
\in SU(n) \}. 
$$ 
However we cannot give a Lie Algebraic condition for (\ref{natext}) 
(which would be an
extension of (\ref{micond}) for this case). Notice that (\ref{micond}) is
essentially the equality of the tangent spaces at $S$ of the two
manifolds in (\ref{orbite2}). The main difficulty is that the two sets
in (\ref{natext}) are not  guaranteed to be manifolds. For example,
if we consider 
\be{ESSE}
S:=\pmatrix{1 & 0 \cr 0 & -1}
\ee 
and $e^{\cal L}:=SO(2)$, and $k=2$, then we have  
\be{insieme}
\{X_1^* {\cal P}(X_2^* S X_2) X_1 |X_1, X_2 \in e^{\cal L}\}=
\{ \pmatrix{a & b \cr b & -a} | a,b \in \RR, \sqrt{a^2+b^2} \leq 1 \}, 
\ee
which is a manifold with boundary.

\vs

Like  for the case of indistinguishability in $1$ step, we can write  
\be{inicon}
\rho(t)=\rho_1(t)+\rho_2(t), 
\ee
with $\rho_1(t) \in {\cal V}_k$ and $\rho_2(t) \in {\cal V}^\perp_k$,
which satisfy  the equations (\ref{OS1}), (\ref{OS2}),
(\ref{OS3}). Therefore if we are interested in the effect on the
output $S$ we can parametrize only the component of $\rho$ in ${\cal
V}_k$. In particular, if ${\cal V}_k$ is the largest of the observability
spaces we can neglect the component $\rho_2(t)$ of the state since it
will not have any effect on any measurement.

\vs 

\noindent{\bf Remark:} The above treatment,  
which has been presented for
Von Neumann measurements,  can be extended to more general types of
measurements (see e.g. \cite{BK},  \cite{BLM}, \cite{BP}, \cite{Davies},
 \cite{gottfried}). We have used the fact that, 
according to the measurement postulate, 
 the state changes as $\rho \rightarrow {\cal
P}(\rho)$. For a more general measurement, 
 with a countable set  of
possible outcomes $\cal M$, 
the state   will change according to 
\be{GC}
\rho \rightarrow {\cal F}(\rho):=\sum_{m \in {\cal M}}\Phi_m(\rho).    
\ee 
The super-operators $\Phi_m$ are called {\it operations} and according
to Kraus' representation theorem \cite{kraus}, under suitable
 assumptions, can be expressed as 
\be{Kraus}
\Phi_m(\rho):=\sum_k \Omega_{mk} \rho \Omega_{mk}^*,  
\ee
for a countable set of operators $\Omega_{mk}$. Our treatment will go
through by replacing ${\cal P}(\rho)$ with ${\cal F}(\rho)$. In
particular,  we can define a dual super-operator ${\cal F}^*$
 acting on observables 
 as 
\be{GC2}
{\cal F}^*(S):=\sum_{m \in {\cal M}}\Phi_m^*(S), \quad \Phi_m^*(S):=
\sum_k \Omega_{mk}^* S \Omega_{mk}.  
\ee 
This has the property $Tr({\cal F}^*(S)\rho)=Tr(S {\cal F}(\rho))$ and
we can use this to extend the calculations in Theorem 5. Moreover the
definition of ${\cal V}_k$ in (\ref{recursspacek}) has to be replaced by 
\be{Vk}
{\cal V}_k:=\oplus_{j=0}^\infty ad_{\cal L}^j {\cal F}^*({\cal
V}_{k-1}). 
\ee 

\section{Initial State Determination}

We now investigate how much information on the initial state we can
extract from an experiment which alternates prescribed evolutions with
measurements. We deal with a single experiment and with a single quantum
 system rather than with many copies  of the same system,  as it is done
some times in this context.  We shall assume,  for simplicity, that
 the system is controllable namely ${\cal L}=su(n)$. Moreover,  we can
assume, without loss of generality,  that the output matrix $S$ is
diagonal. We shall use the following formula (see (\ref{alternyk}),
(\ref{equalk})) for the output at the $k$-th measurement 
\be{USE}
y_k=Tr(X_1 \rho_0 X_1^* {\cal P}
(X_2^* {\cal P}(\cdot \cdot \cdot {\cal P}(X_{k-1}^* {\cal
P}( X_k^* S X_k) X_{k-1}) \cdot \cdot \cdot) X_2)), 
\ee
for the unknown initial state $\rho_0$. 
Now, since every matrix of the type 
${\cal P}(\cdot )$ is diagonal, it 
follows from (\ref{USE}) that it is 
only possible to obtain information
on the diagonal elements of $X_1 \rho_0 X_1^*$ and therefore on at most
$n-1$ independent parameters of the unknown matrix $\rho_0$. 
It is in fact possible to obtain {\it all} the $n-1$
independent diagonal elements of 
the matrix $X_1 \rho_0 X_1^* := \tilde
\rho_0$. At the first measurement 
we obtain 
\be{firstmeas2}
y_1= Tr(\tilde \rho_0 S). 
\ee
Then we choose $X_2$ as a permutation matrix so that $S_2:=X^*_2SX_2$ is
still diagonal but the diagonal elements are a permutation of the
diagonal elements of $S$. We also have ${\cal P}(S_2)=S_2$ so that,  at
the second measurement,  we obtain 
\be{secmeas}
y_2= Tr(\tilde \rho_0 S_2). 
\ee
Then we choose the evolution $X_3$ with $X_3:= \bar X_3 X_2^*$ and 
$\bar X_3$ performing another permutation of the diagonal elements of
$S$. $\bar X_3^* S \bar X_3:= S_3$. Therefore,  the third measurement gives 
\be{thirdmeas}
y_3=Tr(\tilde \rho_0 S_3).  
\ee
Continuing this way, we can 
obtain $n!$ equations for the diagonal
elements of $\tilde \rho_0$, 
$x_1,...,x_n$, i.e. 
\be{linearequationes}
\sum_{k=1}^n a_{jk} x_k=y_j, \qquad j=1,...,n!,  
\ee 
where the elements $a_{jk}$ are appropriate permutations of the diagonal
elements of $S$. 
To this we have to add the equation\footnote{Recall that, 
 without loss of generality,  we are considering  density 
matrices with trace equal to zero rather than one} 
\be{condizioness}
Tr(\tilde \rho_0)=\sum_{k=1}^n x_k =0.  
\ee 
If $S$ is not a scalar matrix, it is always possible to choose $n-1$
permutations and therefore $n-1$ equations in (\ref{linearequationes}) that
together with (\ref{condizioness}) have a unique solution. In fact the 
$n!+1 \times n$ matrix obtained by placing in the first $n!$ rows all
the permutations of the diagonal elements of $S$ and in the last row 
$1,1,...,1$ has always rank $n$. The rank of this matrix is the same as
the rank of a matrix obtained by adding to every row the last row
($1,1,...,1$) multiplied by an arbitrary constant. Therefore we can
assume that the elements of the matrix are nonnegative and apply a Lemma
in Appendix B.

\vs

As seen above, in the Von Neumann case, 
the number of independent parameters that can be inferred
by a sequence of evolutions and measurements is bounded by the dimension
of the system. This suggests to consider different types of measurements
to obtain complete information on the initial state of the system.  One
possible scheme is as follows.  Consider a system
$\Sigma_1$ of dimension $n$, 
with unknown state $\rho_1$ and couple it with a (large)
system $\Sigma_2$, of dimension $m$,  whose state is known 
to be $\rho_2$. The density matrix of the coupled 
system $\rho$ at time $0$ is 
\be{DMtot}
\rho(0)=\rho_1 \otimes \rho_2. 
\ee
This matrix has dimension $nm$ and only $n^2-1$ parameters are
 not known. Now, if we let $\rho$ evolve, after 
time $t$,  the matrix $\rho(t)$ cannot in general 
be written as a tensor product, since the two systems are now 
{\it entangled} \cite{QIC}. If we perform repeated Von Neumann
measurement  
on the coupled system, we are able to obtain information on $nm-1$
independent parameters of $\rho$.  Since $\rho$ contains $n^2-1$ unknown
parameters only, we may be able to  obtain information 
on all of them if $m \geq n$. We give now a simple numerical 
example of this scheme.

\vs

The unknown state of a spin $\frac{1}{2}$ particle is represented by the
 density matrix (without shift of the trace) 
\be{ro1}
\rho_1=\pmatrix{m & l \cr
l^* & 1-m}, 
\ee
with $m$ real. Two  spin $\frac{1}{2}$ particles with known state 
\be{ro2}
\rho_2=\pmatrix{\frac{1}{3} & 0 \cr 0 & \frac{2}{3}} 
\otimes \pmatrix{\frac{1}{3} & 0 \cr 0 & \frac{2}{3}}, 
\ee  
 are  coupled with it. Therefore the unknown state 
\be{uS}
\rho_0:=\rho_1 \otimes \rho_2 
\ee
 has only three
unknown parameters. We can observe the magnetization in the $z$
direction of the system of two spins. The associated 
matrix is given by 
\be{associated}
S=\sigma_z \otimes {\bf 1} \otimes {\bf 1}  + {\bf 1} \otimes \sigma_z 
\otimes {\bf 1} + {\bf 1} \otimes {\bf 1} \otimes \sigma_z 
\ee
(see (\ref{Pauli})) which is diagonal. 
From formula (\ref{USE}) and the
previous discussion we can obtain 
the diagonal elements of the matrix 
$X_1 \rho_0 X_1^*$. 
Consider the  vectors 
\be{vettori}
e_1:=\pmatrix{1 \cr 0} \quad 
e_2:=\pmatrix{0 \cr 1} \quad
v_1:=\frac{1}{\sqrt{2}} \pmatrix{1 \cr 1} \quad
w_1:=\frac{1}{\sqrt{2}} \pmatrix{1 \cr -i}
\ee
If the first three columns of $X_1^*$ are chosen as 
\begin{eqnarray}
\vec x_1:=e_1 \otimes e_1 \otimes e_1 \\
\vec x_2:= v_1 \otimes e_1 \otimes e_2 \nonumber \\
\vec x_3:= w_1 \otimes e_2\otimes e_1) \nonumber 
\end{eqnarray} 
then we obtain for the diagonal elements
\begin{eqnarray}
\vec x^*_1 \rho_0 \vec x_1=\frac{1}{9} m \\
\vec x^*_2 \rho_0 \vec x_2=\frac{1}{9}(1+2 Re(l)) \nonumber \\
\vec x^*_3 \rho_0 \vec x_3=\frac{1}{9}(1+2 Im(l)). \nonumber   
\end{eqnarray}
From this  we can extract the values of $m$ and $l$. 

\section{Discussion and Conclusion}
In this paper,  we have  presented a 
treatment of the observability properties of
quantum systems compatible with quantum  measurement theory. We have
focused on Von Neumann measurements but indicated extensions to more
general types of measurements. 
We have given a characterization of states that cannot be
distinguished in one or more measurements and conditions for
observability. 
Contrary to most studies  on observability of nonlinear
systems (see e.g.  \cite{Sontag}) 
conditions  for observability and indistinguishability are {\it  global}
in this case, however observability does not always imply that it is possible
to infer from appropriate evolutions and measurements all the parameters
of the initial state. In fact, for Von Neumann measurements, 
 there is a natural limit to the number of
parameters of the state that can be 
derived. This   does not improve if we consider
measurements of different (and not necessarily commuting)
observables. In this case, the result of the $k-$th measurement has
still the form (\ref{USE}) although now each projection $\cal P$
corresponds to a possibly different observable   and $S$
corresponds to the observable measured last. In 
this case the first $\cal P$ on the left 
is {\it always} the projection 
corresponding to the first measurement 
and therefore, once again, only
at most $n-1$ independent parameters 
of $X_1 \rho_0 X_1^*$ can be obtained. We have seen, in the previous
section,  that complete  information on the initial state may 
 be obtained by coupling  the system with an auxiliary system whose state is known. 


\vs

\noindent{\bf Acknowledgment} 
This research was supported by NSF under Career Grant ECS-0237925.

\section*{Appendix A: Evaluation of $\cal V$ using a set of generators
of $\cal L$. }

Let $B_1,...,B_m$ a set of generators of $\cal L$. Denote by 
$\bar {\cal V}$ the space spanned by the matrices in (\ref{mat}). 
It is obvious that 
\be{obvious}
\bar {\cal V} \subseteq \oplus_{k=0}^\infty ad_{\cal L}^k iS. 
\ee
To show 
\be{notobvious}
 \oplus_{k=0}^\infty ad_{\cal L}^k iS \subseteq \bar {\cal V},  
\ee
we first show that 
\be{inclus}
[\bar {\cal V}, {\cal L}] \subseteq \bar {\cal V}. 
\ee
It is enough to show for elements $F$ in a 
 basis of $\cal L$ given by
 $B_1,...,B_m$, and linearly independent (repeated)  Lie brackets, 
$[F, \bar {\cal V}] \subseteq {\bar {\cal V}}$. We proceed by induction on
the depth of $F$. If $F$ is of depth $0$, namely $F$ is one of the
matrices $B_1,...,B_m$ then (\ref{inclus}) follows from the
definition of $\bar {\cal V}$. Now,  let us assume (\ref{inclus}) true 
for matrices $F$ of depth $\leq d$ and let us show it for matrices $F$ of
depth $d+1$. In particular, write $F$ as $F:=[Z,T]$, where $Z$ is of
depth $d$ and $T$ is of depth zero. If $\bar V$ is a matrix in $\cal V$, from
the Jacobi identity,  we obtain 
\be{Jacobi}
[V,[Z,T]]=-[Z,[T,V]]-[T,[V,Z]], 
\ee  
since both terms on the right hand side are in $\bar {\cal V}$, 
from the inductive
assumption, we have that the term on the left hand side is also 
in $\bar {\cal V}$, therefore we have proved (\ref{inclus}). Now, from
(\ref{inclus}) we have 
\be{jj}
ad_{\cal L}iS:= [iS, {\cal L}] \subseteq \bar {\cal V}, 
\ee  
and from this 
\be{jjj}
ad_{\cal L}^2iS:=[[iS, {\cal L}],{\cal L}] 
\subseteq [\bar {\cal V}, {\cal L}] \subseteq \bar {\cal V}, 
\ee
where we have used (\ref{inclus}). Proceeding 
this way, we see that for 
every $k \geq 0 $ 
\be{inlus2}
ad_{\cal L}^kiS \subseteq \bar {\cal V}, 
\ee
which proves (\ref{notobvious}).

\section*{Appendix B}

{\bf Lemma } {\it 
Let $x_{1}, \ldots,x_{n}$ be $n$ non-negative numbers not all equal. 
 Consider the  matrix 
$A(x_{1},\ldots,x_{n})\in \RR^{n!\times n}$ whose rows are the
 the permutations of $x_1,...,x_n$. Then 
the matrix $A=A(x_{1},\ldots,x_{n})$ has  rank $n$.}

{\bf Proof}

Let  $2\leq r\leq n$ be the number of different values assumed by
the $\{x_i\}_{i=1,\ldots,n}$. Denote by $0\leq d_{1}<\cdots <
d_{r}$ these values, and let $l_i$, for $i=1,\ldots,r$ be the
cardinality of $\{ \, j\, |\, x_j=d_i\, \}$.  Thus
$\sum_{i=1}^{r}l_{i}=n$. We will prove our  statement on induction
on $r\geq 2$.

{\bf{case $r=2$}}

We prove this part by induction on $n\geq 2$. If $n=2$, then the
statement is easily proved by computing the determinant of $A$.
Let $n>2$. Since all the columns of $A$ sum up to the same value,
which is strictly positive, setting 
\be{somma1}
 A'=   \left(
\begin{array}{c}
                   1,\ldots, 1 \\
           A \end{array} \right),   
\ee
we have that \be{somma} \text{rank} A
= \text{rank} A'= \text{rank}  \left(
\begin{array}{c}
                   1,\ldots, 1 \\
           A \end{array} \right), 
           \ee
where we have set 

Choose a value ${x_{\bar{i}}}\in \{ x_{1},\ldots,x_{n} \}$, such
that ${x_{\bar i}}=d_{1}$, assuming that $l_1 \geq 2$ (otherwise choose
it so that ${x_{\bar i}}=d_{2}$). Assume that we have rearranged the
rows of $A'$ in such a way that the first element of the second to
the $(n-1)!+1$-th row  is $x_{\bar{i}}$. Notice that this can be done
since the rank remains unchanged. Then  for $i=2,\ldots,(n-1)!+1$ we
subtract from the $i$-row of $A'$  the first row multiplied by
$x_{\bar{i}}$. Notice that if $d_{1}=0$ we leave the matrix
unchanged. After this operation, the matrix $A'$ has the following
form: \be{form1} A'= \left(
\begin{array}{c}
         \begin{array}{lc}
         1 &  1, \cdots, 1 \\
         0 &   \\
         \vdots   & \tilde{A} \\
         0 &  \\
        \end{array}
         \\
              B
          \end{array} \right).
          \ee

Notice that $\tilde {A}$ is an $(n-1)!\times (n-1)$-matrix with the same
structure of  $A$ and  values $y_{j}=
x_{j}-x_{\bar{i}}\geq 0$, in particular the $y_j$'s  are
either $0$ or $d_2-d_1$. Thus, by inductive assumption we have
that $\text{rank} \tilde{A}= n-1$, which, in turn, implies
$\text{rank} A' =n$ as desired. Had we chosen 
$x_{\bar i}=d_2$, we would have had all the values $y_j \leq 0$ with the
two possible values $0$ and $d_1-d_2$, then we would have changed the
sign of $\tilde A$ (which does not affect the rank) 
and applied the inductive assumption.  

{\bf{case $r>2$}}
We assume that the result is true for $r-1$. 
The idea of the proof is similar to the $r=2$ case. Assume again
that we have chosen   ${x_{\bar{i}}}\in \{ x_{1},\ldots,x_{n} \}$,
such that ${x_{\bar i}}=d_{1}$ and we have performed to the matrix
$A'$ (defined in (\ref{somma1})) the same operation as in the
previous case to put $A'$ in the form (\ref{form1}).

Now, if we prove that $\text{rank} \tilde{A}= n-1$ then we get
$\text{rank} A' =n$. As before,  $\tilde {A}$ is an $(n-1)!\times
(n-1)$-matrix which is the same structure as   $A$ with values
$y_{j}= x_{j}-x_{\bar{i}}\geq 0$ for $j\neq \bar{i}$. If
$l_{1}=1$, then we are done by the inductive assumption since the
numbers $y_{j}$  assume $r-1$ different nonnegative values. If $l_{1}>1$, then
we perform the same procedure as before starting with $\tilde{A}$
instead of $A$. Notice that $\tilde{A}$ has  $n-1$ different 
numbers, and is such that $d_1=0$ and the cardinality of the
$\{y_j=0\}$ is $l_1-1$. Thus we need to repeat this procedure $l_1$
times and then we can  conclude by induction.

\qed

\end{document}